\documentclass[onecolumn,showpacs,preprintnumbers,amsmath,amssymb,nofootinbib,APS]{revtex4}

\usepackage{dcolumn}
\usepackage{bm}
\usepackage[latin1]{inputenc}
\usepackage[spanish,english]{babel}
\usepackage{amsfonts}
\usepackage{amssymb}
\usepackage{graphicx}

\newcommand{\be}{\begin{equation}}
\newcommand{\ee}{\end{equation}}
\newcommand{\bea}{\begin{eqnarray}}
\newcommand{\eea}{\end{eqnarray}}

\begin{document}
\preprint{hep-th/}

\title{{\bf Acceleration radiation and the Planck scale}}
\author{Iván Agulló}\email{ivan.agullo@uv.es}
\affiliation{ {\footnotesize Department of Physics, University of
Maryland, College Park, Maryland 20742 \\and  \\ Departamento de
Física Teórica and IFIC, Centro Mixto Universidad de
Valencia-CSIC.
    Facultad de Física, Universidad de Valencia,
        Burjassot-46100, Valencia, Spain. }}
\author{José Navarro-Salas}\email{jnavarro@ific.uv.es}
\affiliation{ {\footnotesize Departamento de Física Teórica and
IFIC, Centro Mixto Universidad de Valencia-CSIC.
    Facultad de Física, Universidad de Valencia,
        Burjassot-46100, Valencia, Spain. }}

\author{Gonzalo J. Olmo}\email{golmo@perimeterinstitute.ca}
\affiliation{ {\footnotesize Perimeter Institute, 31 Caroline St. N, Waterloo, ON N2L 2Y5, Canada \\ and \\ Physics Department, University of
Wisconsin-Milwaukee, P.O.Box 413, Milwaukee, WI 53201 USA}}

\author{Leonard Parker}\email{leonard@uwm.edu}
\affiliation{ {\footnotesize Physics Department, University of
Wisconsin-Milwaukee, P.O.Box 413, Milwaukee, WI 53201 USA}}

\date{February 28th, 2008}

\begin{abstract}
A uniformly accelerating observer perceives  the Minkowski vacuum
state as a thermal bath of radiation. We  point out that this
field-theory effect can be derived, for any dimension higher than
two, without actually invoking very high energy physics. This
supports the view that this phenomenon is robust against
Planck-scale physics and, therefore, should be compatible with any
underlying microscopic theory.

\end{abstract}


\maketitle

\section{Introduction}

The fact that the notion of particles is ambiguous in a general
curved spacetime plays a crucial role to derive gravitational
particle production. In cosmology this was first exploited in
\cite{parker}, and for black holes in \cite{hawk1}. The expansion of
a field in two different sets of positive frequency modes:
$u^{in}_j(x)$ (usually defined at past infinity) and $u^{out}_j(x)$
(defined at future infinity) leads to a relation for the
corresponding creation and annihilation operators: $
a_i^{out}=\sum_j (\alpha^* _{ij}a_j^{in}-\beta ^*_{ij}a_j^{in
\dagger})$. When the coefficients $\beta _{ij}$ do not vanish  the
vacuum states $|in\rangle$ and $|out\rangle$ do not coincide and,
therefore, the number of particles measured in the $i^{th}$ mode by
an ``out'' observer  in the state $|in\rangle$ is given by
 $\langle in| N_i^{out}|in \rangle = \sum_k
|\beta_{ik}|^2$. This general framework leads to two important
predictions: particle creation in an expanding universe and in a
 gravitational collapse.

 However, even in Minkowski space,
the existence of two inequivalent quantizations leading to different concepts of particles was first
pointed out in \cite{fulling}, continued in \cite{davies}, and
crucially understood in terms of particle detectors in
\cite{unruh1}. In short, the standard Minkowski vacuum state is
perceived by an accelerated observer as a thermal bath of
particles at the temperature $T= \frac{\hbar a}{2\pi c k_B}$,
where $a$ is the acceleration. This effect shares some physical
and mathematical aspects with the one discovered by Hawking (for a
complete account see \cite{waldbook}). However, they are indeed
distinct. For instance, in the Hawking effect there is an outgoing
thermal energy flux at future infinity, as perceived by an
inertial observer there. In contrast, in the acceleration
radiation there is no net energy flux at infinity. Only a thermal bath of
radiation exists for the uniformly accelerated observer.

On the other hand, the derivation of the Hawking effect seems to
invoke Planck-scale physics (see, for instance
\cite{hooft-jacobson}). Any out-going Hawking quanta will have an
exponentially increasing frequency when propagated backwards in
time and measured by a free-falling observer. Accordingly, any
microscopic structure of a quantum gravity theory could leave some
imprint or signal in the spectrum of radiation. However, the
results of string theory agree with Hawking's prediction (for low
emission frequencies) \cite{strings}. The string theory
microscopic description of black hole emission can also be
extended to the super-radiance regime of some extremal rotating
black holes \cite{dias-emparan-maccarrone}, which can be obtained
at the semiclassical level as a limiting case $T \to 0$ of the
general Hawking radiation formula \cite{hawk1}. All the above
suggests that a quantum gravity theory, with new degrees of
freedom at the Planck scale, should not necessarily modify the
bulk of the semiclassical effects, at least for low emission
frequencies. This is essentially the conclusion of the
investigation of \cite{agullo-navarro-salas-olmo-parker}, which
argued, within a purely field theory framework, that ultrashort
distances do not significatively contribute to the Hawking effect,
if Lorentz invariance is somehow respected. Only at high emission
frequencies could an underlying theory of quantum gravity
potentially predict important deviations from semiclassical
results.

In \cite{agullo-navarro-salas-olmo-parker} we derived
(gravitational) particle production rates in terms of the
two-point function of the matter field.  Within this new approach
we have studied the role of Planck scale in the spectrum of
Hawking radiation. The aim of this paper is to extend our analysis
focusing now on the acceleration radiation. We find that the acceleration radiation
is more  insensitive to trans-Planckian physics than is the Hawking radiation,
at least in four dimensions.

In Section II we briefly review the basics of the acceleration radiation. In Section III
we show how  the acceleration radiation (Fulling-Davies-Unruh effect) can be obtained easily in terms of
correlation functions. Section IV is devoted to a discussion of the role of Lorentz
invariance in generating the Planckian spectrum. We point out that
this  effect can be derived, for any dimension higher than two,
without actually invoking very high energy physics. This suggests
that this phenomenon is robust against Planck-scale physics and,
therefore,  that it should persist when microscopic degrees of
freedom are considered.

\section{Acceleration radiation}

 We shall present in
this section the main physical aspects of the standard derivation
of the acceleration radiation.  Let us consider an observer with
a uniformly accelerated trajectory (from now on we take $c=1$),
known as a Rindler observer,
\begin{equation} \label{accelerated}
T =\frac{e^{a\xi}}{a}
\sinh{at} \ , \
X=\frac{e^{a\xi}}{a} \cosh{at} \ , \ Y=Y_0 \ , \ Z=Z_0
\end{equation}
and a massless scalar field propagating in the Minkowskian
background spacetime. The wave equation  $\Box \phi(x)=0$ in the
coordinates of the accelerated observer becomes \be
(e^{-2a\xi}(-\partial^2_t+\partial^2_\xi)+\partial^2_Y+\partial^2_Z
)\phi(t,\xi,Y,Z)=0\ee The $Y,Z$ dependence can be trivially
integrated using  plane waves $\phi(t,\xi,Y,Z)=\phi(t,\xi) e^{i k_Y
Y} e^{i k_Z Z}$. Introducing this ansatz in the equation, we find
\be \label{eq:waveq}
[(-\partial^2_t+\partial^2_{\xi})-e^{2a\xi}(k_Y^2+k_Z^2)]\phi(t,\xi)=0
\ee This equation indicates that the free scalar field of the
Minkowski observer appears like a scalar field in a repulsive
potential $V(\xi) \propto e^{2a\xi}\vec{k}_{\bot }^2$, where
$\vec{k}_{\bot}^2 = k_Y^2+k_Z^2$, for the uniformly accelerated
observer. The exact form of the normalized modes, with natural
support on the accessible region for the accelerated observer
(right-hand Rindler wedge), can be expressed as
\begin{equation}
u^{R}_{w,\vec{k}_{\bot}}=\frac{e^{-iwt}}{2\pi^2\sqrt{a}}\sinh^{\frac{1}{2}}\left(\frac{\pi
w}{a}\right)K_{iw/a}\left(\frac{|\vec{k}_{\bot
}|}{a\hbar}e^{a\xi}\right)e^{i\vec{k}_{\bot}\cdot\vec{X}_{\bot}}
\end{equation}
The important point is that the above positive frequency modes
cannot be expanded in terms of the standard purely positive
frequency modes of the inertial observer \be
u^{M}_{k_X,\vec{k}_{\bot}}
=\frac{1}{\sqrt{2(2\pi)^3k_0}}e^{-ik_0T+i(k_X X +
\vec{k}_{\bot}\cdot\vec{X}_{\bot})} \ , \ee where $k_0= \sqrt{k_X^2
+ \vec{k}_{\bot}^2}.$ The detailed analysis requires one to compute
the corresponding Bogolubov coefficients. They are found to be
\cite{fulling,crispino} \be \beta_{w \vec{k}_{\bot}, k'_X
\vec{k}'_{\bot}}=-\left [2\pi ak'_0(e^{2\pi w/a}-1)\right ]^{-1/2}
\left (\frac{k'_0 + k'_X}{k'_0 - k'_X} \right )^{-iw/2a}\delta
(\vec{k}_{\bot}-\vec{k}'_{\bot}) \ . \ee The mean number of Rindler
particles in the Minkowski vacuum is obtained as the integral \be
\int_{-\infty}^{+\infty}d\vec{k}' \beta_{w_1 \vec{k}_{\bot},
\vec{k}'} \beta_{w_2 \vec{k}_{\bot}, \vec{k}'}^* \ . \ee The
integration in $k'_X$ reduces to \be
\label{integralkX}\int_{-\infty}^{+\infty}dk'_X (2\pi ak'_0)^{-1}
\left (\frac{k'_0 + k'_X}{k'_0 - k'_X} \right
)^{-i(w_1-w_2)/2a}=\delta (w_1 - w_2) \ , \ee and taking into
account the remaining terms one easily gets  \be
\int_{-\infty}^{+\infty}d\vec{k}' \beta_{w_1 \vec{k}_{\bot 1},
\vec{k}'} \beta_{w_2 \vec{k}_{\bot 2}, \vec{k}'}^* =
\frac{1}{e^{2\pi w_1/a}-1} \delta (w_1 - w_2) \delta (\vec{k}_{\bot
1}-\vec{k}_{\bot 2})\ . \ee The final outcome becomes then extremely
simple. A uniformly accelerated observer feels himself immersed in a
thermal bath of radiation at temperature $ k_BT = a \hbar/2\pi$.

This result is reinforced by Unruh's operationalism interpretation
\cite{unruh1}. In short, the particle content of the vacuum
perceived by an accelerated observer with motion $x=x(\tau)$ can
be described by the  response function $F(w)$ of an ideal quantum
mechanical detector (see also \cite{birrel-davies}) \be F(w)=
\int_{-\infty}^{+\infty}d\tau_1\int_{-\infty}^{+\infty}d\tau_2
e^{-iw(\tau_1 -\tau_2)}\langle \phi(x(\tau_1))\phi(x(\tau_2))
\rangle \ ,  \ee where \be \label{Wfunction}\langle
\phi(x_1)\phi(x_2) \rangle = \frac{\hbar}{4\pi^2 (-(T_1 - T_2
-i\epsilon)^2 + (X_1 - X_2)^2 + (Y_1 - Y_2)^2+ (Z_1 - Z_2)^2)}\ ,
\ee is the two-point function of the field evaluated in the
Minkowski vacuum. For a uniformly accelerated trajectory the
above response function, or better, the rate ${\dot F(w)}$ turns
out to be \be \label{rateF}{\dot F(w)}=
\int_{-\infty}^{+\infty}d\Delta \tau e^{-iw\Delta \tau}\langle
\phi(x(\tau_1))\phi(x(\tau_2)) \rangle \ . \ee Performing the
integral one obtains \be {\dot F(w)}=
\frac{1}{2\pi}\frac{\hbar w}{e^{2\pi w/a} -1} \ . \ee

It is important to note that in either approach the thermal spectrum seems to depend
crucially on the validity of relativistic field theory on all
scales. In the former, the intermediate
integral (\ref{integralkX}) involves an unbounded integration
in arbitrary large Minkowskian momentum $k'_X$. If one introduces
an ultraviolet cutoff $\Lambda$ for $|k'_X|$ in the above integral, which particularizes a given Lorentz frame,
the resulting thermal spectrum is largely truncated. In the
detector model approach, the role of high energy scale emerges in
the evaluation of the integral (\ref{rateF}), which crucially
depends on the short-distance behavior of the Wightman function
(\ref{Wfunction}).

\section{Acceleration radiation and two-point functions}

We will now present the formalism used in \cite{agullo-navarro-salas-olmo-parker} and then will apply it to the calculation of the acceleration radiation.

\subsection{Particle creation and two-point functions}

 Let us suppose that $\phi$ is a scalar field propagating in an arbitrary spacetime. We can rewrite the
expectation values of the operator $N_{i}^{out}\equiv \hbar^{-1}\
{a^{out}}^\dagger_{i} a^{out}_{i} $, in terms of the corresponding
scalar product for the field \bea \label{Nijout} &&\langle
in|N_{i}^{out}|in\rangle =\sum_k \beta_{ik}\beta_{ik}^*=-\sum_k
(u^{out}_{i}, u^{in*}_k)(u^{out*}_{i}, u^{in}_k)=\nonumber \\
&=&\sum_k\left(\int_{\Sigma}d\Sigma^{\mu}_1u^{out}_{i}(x_1)
{\buildrel\leftrightarrow\over{\partial}}_\mu u^{in}_k(x_1)\right)
\left(\int_{\Sigma}d\Sigma^{\nu}_2u^{out
*}_{i}(x_2){\buildrel\leftrightarrow\over{\partial}}_\nu u^{in
*}_k(x_2)\right)\ , \ \eea where $\Sigma$ is an initial Cauchy
hypersurface. If we now consider the sum of {\it in} modes before making
the integrals of the two scalar products, and take into account
that \be \langle in| \phi (x_1)\phi (x_2)| in \rangle=\hbar\sum_k
u_k^{in}(x_1){u_k^{in}}^*(x_2) \ , \ee we obtain a simple
expression for the particle production number in terms of the
two-point function
\begin{equation}\label{eq:N-eps}
\langle in|N_{i}^{out}|in\rangle = \hbar^{-1} \int_\Sigma
d\Sigma_1 ^\mu d\Sigma_2 ^\nu
[u^{out}_{i}(x_1){\buildrel\leftrightarrow\over{\partial}}_\mu
][u^{out*}_{i}(x_2){\buildrel\leftrightarrow\over{\partial}}_\nu
]\langle in| \phi (x_1)\phi (x_2)|in\rangle \ .
\end{equation}
The above expression requires to interpret  the two-point function
 in the
distributional sense. The ``$i\epsilon$-prescription''  and the
Hadamard condition\footnote{The two-point distribution   should
have (for all physical states) a short-distance structure similar
to that of the ordinary vacuum state in Minkowski space:
$(2\pi)^{-2}(\sigma +2i\epsilon t + \epsilon^2)^{-1}$, where
$\sigma(x_1,x_2)$ is the squared geodesic distance
\cite{waldbook}.}  should be assumed for $\langle in| \phi
(x_1)\phi (x_2)|in\rangle$, as in (\ref{Wfunction}). However,
taking into account the trivial identity $\langle
out|{a^{out}}^\dagger_{i} a^{out}_{i}|out\rangle =0$ we can
rewrite the above expression as
\begin{eqnarray} \label{eq:N-nord}
&&\langle in|N_i^{out}|in\rangle = \frac{1}{\hbar} \int_\Sigma
d\Sigma_1 ^\mu d\Sigma_2 ^\nu
[u^{out}_{i}(x_1){\buildrel\leftrightarrow\over{\partial}}_\mu
][u^{out*}_{i}(x_2){\buildrel\leftrightarrow\over{\partial}}_\nu ]
\nonumber \\ && \times[\langle in| \phi (x_1)\phi (x_2)|in\rangle
- \langle out| \phi (x_1)\phi (x_2)|out\rangle] \ .
\end{eqnarray}
Now the Hadamard condition for both $|in\rangle$ and
$|out\rangle$ states ensures that the difference of the above
two-point distributions is a smooth function.

 Intuitively the idea behind the above manipulations is simple. In the
conventional analysis in terms of Bogolubov coefficients, we first
perform the integration in distances  and leave to the end the sum
of {\it in} modes. In contrast, we can invert the order and perform
first the sum of {\it in} modes, which naturally leads to introduce the
two-point function of the matter field, and leave the integration
in distances to the end. Despite this simple technicallity, one should
not underestimate the physical content of expression (\ref{eq:N-nord}).
The existence of different correlations $\langle  \phi (x_1)\phi (x_2)\rangle $
between {\it in} and {\it out} observers, weighted by the form of the
modes of the detected quanta, is at the root of the phenomenon of
particle production. Moreover,  the relevant correlations are
those with support in the region where the wave-packet modes are
peaked.

One of the advantages of expression (\ref{eq:N-nord}), as compared
 with (\ref{eq:N-eps}),  is that it displays clearly the possible symmetries
of the problem. For instance, for a conformal field theory and for
{\it in} and {\it out} modes  related by spacetime conformal
transformations, the integrand (\ref{eq:N-nord}) is manifestly
zero. In contrast, it is the full integral (\ref{eq:N-eps}) which
vanishes.

The ``$i\epsilon$-prescription'' described above, when applied to
a gravitational collapse turns out to be somewhat parallel  to the
approach of \cite{fredenhagen-haag90}. Here, as in
\cite{agullo-navarro-salas-olmo-parker}, we  want to put forward
expression (\ref{eq:N-nord}) to evaluate the particle production
rate and to analyze the role of the Planck scale.

\subsection{Rederiving the acceleration radiation}

 We shall now
explain with some detail how the Fulling-Davies-Unruh effect can
be derived using the two-point functions of the field. We denote
the Rindler modes (``out'' in the above notation) by
$u_i^{R}=\phi_w(t,\xi)e^{- i\vec{k}_{\bot}\vec{x}_{\bot }} $,
where $i\equiv(w,\vec{k}_{\bot })$, and the Rindler vacuum by $|0_R\rangle$ ( $|out\rangle$ in the above notation). The Minkowski vacuum will be denoted by $|0_M\rangle$ ($|in\rangle$ in the above notation). Now that the notation of this section has been fixed,  we will explain how to choose a suitable Cauchy hypersurface to evaluate the integrals of (\ref{eq:N-nord}).
To compute the number of
particles for the accelerated observer  in the Minkowski vacuum
$|0_M\rangle$, one can
naturally choose a hyperplane $T - \lambda X= constant$, with
$|\lambda | <1$, as the initial Cauchy surface. However, it is
convenient to consider the limiting case $\lambda=1$ and the null
plane $H^+_0 $, defined as $U\equiv T-X=U_0<0$ (or the analogous
null plane $H^-_0$, defined as $V\equiv T+X=V_0>0$)  as our
initial data hypersurface. As emphasized in \cite{waldbook}
(section 5.1), any solution of the massless Klein-Gordon equation
in Minkowski space, having any dimension greater than two, is
uniquely determined by its restriction to the hyperplane $H^+_0$
alone (or $H^-_0$ alone). So $H^+_0$ (or $H^-_0$) is enough to
characterize the field configuration\footnote{In two dimensions
$H^+_0$ is not enough and we need $H^+_0 \bigcup H^-_0$ to have a
proper initial surface.} and can be used as the initial value
surface $\Sigma$. Therefore, we convert (\ref{eq:N-nord}) into (we
introduce two indices $i_1$ and $i_2$ since we are using
plane-wave type modes instead of wave-packets) \bea \label{number}
\langle_M 0|{N}_{i_1,i_2}^{R}|0_M \rangle =
\frac{4}{\hbar}\int_{H^+_0}dv_1d\vec{x}_{\bot 1}dv_2d\vec{x}_{\bot
2} u^{R}_{i_1}(x_1) u^{R*}_{i_2}(x_2) \nonumber \\
 \partial_{v_1}\partial_{v_2}[\langle
_M 0|\phi(x_1)\phi(x_2)|0 _M\rangle -\langle _R
0|\phi(x_1)\phi(x_2)|0 _R\rangle ] \ . \eea Since the accelerated
modes $u^{R}_{i}(x)$ have support on the right-handed Rindler
wedge the above integral is naturally restricted to the right
wedge part of $H^+_0$. The relevant derivatives of the two-point
functions in the Minkowski vacuum can be expressed, using the
inertial null coordinates $V,U$, as:
 \bea
\langle_M
0|\partial_{V_1}\phi(x_1)\partial_{V_2}\phi(x_2)|0_M\rangle=
\frac{1}{4 \pi^2} \int d\vec{k}_{\bot } G^M_{\vec{k}_{\bot
}}(x_1,x_2) e^{-i \vec{k}_{\bot } \vec{x}_{\bot }} \ , \eea where,
at the region $H^+_0$: \bea
G^M_{\vec{k}_{\bot}}(x_1,x_2)|_{H_0^+}=
\partial_{V_1}\partial_{V_2} \frac{\hbar}{2 \pi}
K_0(|\vec{k}_{\bot}|\sqrt{-(V_1-V_2)(U_1-U_2)})|_{H^+_0}
=-\frac{\hbar}{4 \pi} \frac{1}{(V_1-V_2)^2}\ ,\eea where
 $K_0(x)$ is a modified Bessel function. Therefore,
 \be
\langle_M0|\partial_{V_1}\phi(x_1)\partial_{V_2}\phi(x_2)|0_M\rangle|_{H^+_0}=
  -\frac{\hbar}{4
\pi} \frac{1}{(V_1-V_2)^2}\delta(\vec{x}_{\bot 1}- \vec{x}_{\bot
2}) \ . \ee It is now convenient to perform the calculation on the
null plane $H^+$, obtained by the limiting case $U_0 \to 0$. As
approaching to $H^+$ ($\xi \to -\infty$), the potential decays
exponentially and the (right)-Rindler modes can be approximated as
\be \label{eq:Rmodes} u^R_i = \frac{e^{i
\gamma(w,|\vec{k}_{\bot}|)}}{(2\pi)^{3/2}\sqrt{2w}}(e^{-iwu}+
re^{-iwv})e^{i \vec{k}_{\bot } \vec{x}_{\bot }} \ , \ee where
$v=t+\xi$,$u=t-\xi$, $e^{i
\gamma(w,|\vec{k}_{\bot}|)}=\Gamma[1+i\frac{w}{a}]^{-1}\left(\frac{|\vec{k}_{\bot}|}{2a}\right)^{i\frac{w}{a}}$,
and $r(w, \vec{k}_{\bot })=e^{-i2\gamma(w,|\vec{k}_{\bot}|)}$ is
the reflection amplitude.

 Moreover, the two-point function in the accelerated
Rindler vacuum can also be worked out as \bea
\langle_R0|\partial_{v_1}\phi(x_1)\partial_{v_2}\phi(x_2)|0_R\rangle|_{H^+}
=\frac{1}{4 \pi^2} \int d\vec{k}_{\bot
}G^R_{\vec{k}_{\bot}}(x_1,x_2)|_{H^+} e^{-i \vec{k}_{\bot }
\vec{x}_{\bot }} = -\frac{\hbar}{4 \pi}
\frac{1}{(v_1-v_2)^2}\delta(\vec{x}_{\bot 1}- \vec{x}_{\bot 2}) \ .
\eea With this, the equation (\ref{number}) becomes:

\bea \label{integralV}\langle_M 0|{N}_{i_1,i_2}^{R}|0_M \rangle =
-\frac{|r|^2}{4\pi^2\sqrt{w_1w_2}}\int_{V_1, V_2 >0}
e^{-iw_1v_1+iw_2v_2}\nonumber
\\\left[\frac{dV_1dV_2}{(V_1-V_2)^2}-\frac{dv_1dv_2}{(v_1-v_2)^2}\right]
\delta(\vec{k}_{\bot 1}-\vec{k}_{\bot 2})\eea where the
transversal $(Y,Z)$ dependence has been trivially integrated,
producing the delta function. We would like to emphasize again the
physical meaning of the above expression. The particle content of
the Minkowski vacuum, as perceived by the accelerated observer, is
displayed as an integral measuring the different vacuum
correlations of inertial and accelerating observers, weighted by
the form of the modes of the accelerated observer.

Taking into account that the relation between the null inertial
coordinate $V$ and the accelerated one $v$ is $V=a^{-1} e^{a v}$
we then have\bea \label{formulaN}& &\langle_M 0|N^R_{i_1,i_2}|0_M
\rangle= - \frac{|r|^2}{4 \pi^2\sqrt{w_1 w_2}}
\int_{-\infty}^{\infty} dv_1 dv_2 e^{-i w_1 v_1} e^{i w_2
 v_2} \left[ \frac{(a/2)^2}{\sinh^2{\frac{a}{2}(v_1-v_2)}}-\frac{1}{(v_1-v_2)^2}\right]\delta(\vec{k}_{\bot 1}-\vec{k}_{\bot 2}) \ . \eea
Note that $|r|=1$ because the reflection amplitude $r$ is a pure phase.  Integrating over $v_1 + v_2$ we are left with (we define
$\Delta v \equiv v_1 - v_2$)  \bea \label{integral3}\langle_M
0|N^R_{i_1,i_2}|0_M \rangle&=& - \frac{1}{4 \pi^2 w_1}
\int_{-\infty}^{\infty} d(\Delta v) \ e^{-i w_1\Delta v} \left[
\frac{(a/2)^2}{\sinh^2{\frac{a}{2}\Delta v}}-\frac{1}{(\Delta
v)^2}\right]\delta
(w_1 - w_2)\delta(\vec{k}_{\bot 1}-\vec{k}_{\bot 2}) \nonumber \\
  &=&\frac{1}{e^{2\pi w_1/a} -1}\delta(\vec{k}_{1}-\vec{k}_{2}) \ . \eea

Alternatively, we can choose  the null hypersurface $H^-$ (defined as $V_0=0$), instead of $H^+$, as our initial hypersurface.
Then one finds \bea \label{integralU}\langle_M
0|{N}_{i_1,i_2}^{R}|0_M \rangle =
-\frac{1}{4\pi^2\sqrt{w_1w_2}}\int_{U_1, U_2 <0}
e^{-iw_1u_1+iw_2u_2}\nonumber
\\\left[\frac{dU_1dU_2}{(U_1-U_2)^2}-\frac{du_1du_2}{(u_1-u_2)^2}\right]
\delta(\vec{k}_{\bot 1}-\vec{k}_{\bot 2}) \ , \eea  where now $U=
-a^{-1} e^{-a u}$. This leads again to the same Planckian
spectrum. \\

Note that, since the Rindler modes (\ref{eq:Rmodes}) have
 a reflection amplitude of unit module,
we get a null net flux of radiation but a non-vanishing thermal
energy density able to excite an accelerated particle detector.
In contrast, because in two spacetime dimensions the left and right
modes are independent and one needs $H^+ \cup H^-$ to have a
complete initial surface, both expressions (\ref{integralV}) and
(\ref{integralU}) would then be needed to properly obtain the acceleration radiation.

\section{Acceleration radiation, Lorentz invariance and the Planck scale}

Let us now examine with more detail the basic formulas
(\ref{integralV}) and (\ref{integralU}) leading to the thermal
spectrum. As remarked above, those formulas tell us that particles
in a given mode stem from the different two-point correlations
seen by inertial and accelerated observers. In this sense, one could think that the
Fulling-Davies-Unruh effect seems to require the validity of
special relativity for arbitrarily large and unbounded boosts. This is so
because the affine distance  $|V_1 - V_2|$ along the null plane
$H^+$ in the direction of the acceleration  can be made arbitrarily small
\be \label{contraction}
(V_1 - V_2)^2 \sim e^{2av_1}(v_1 -v_2)^2 \ , \ee
as  perceived by the accelerated observer as $v_1\approx v_2 \to -\infty$.
As a consequence, even if $|v_1 - v_2|\equiv |\Delta v|$ is well above
the Planck length $l_P$,  $|V_1 - V_2|$ involves
sub-Planckian distances when $v_1\approx v_2 \to -\infty$ due to the
extreme length contraction. And this ultra-high length contraction seems
fundamental in (\ref{integralV}) for getting the exact thermal result.
We will show next, however, that the bulk of the Fulling-Davies-Unruh
effect does not require the consideration of sub-Planckian lengths. \\

Let us first note that we work in a Lorentz invariant framework. Despite this fact, in the evaluation of (\ref{integralV}) we explicitly used a particular inertial observer related to the accelerated observer by the change of coordinates (\ref{accelerated}) [see (\ref{formulaN})]. In a fully Lorentz invariant framework, however, this choice of inertial observer is arbitrary, since the outcome of (\ref{integralV}) is actually independent of the particular inertial observer chosen\footnote{The proof of this claim is simple. We just need to look at the relation between inertial and accelerated observers given in (\ref{accelerated}), and consider another inertial observer related to the first one by $T'=\gamma(T+\beta X), X'=\gamma(X+\beta T)$. The relation between the null inertial coordinates of the new inertial observer and the accelerated observer is given by
$V'\equiv T'+ X'=\gamma(1+\beta)\frac{1}{a}e^{a v}$ and $U'\equiv T'- X'=-\gamma(1-\beta)\frac{1}{a}e^{-a u}$. Using these relations, we can evaluate the terms $dV_1dV_2/(V_1-V_2)^2$ and $dU_1dU_2/(U_1-U_2)^2$ of (\ref{integralV}) and (\ref{integralU}), respectively, and find that all $\beta$ and $\gamma$ factors disappear.}. In such a framework, the only variable that can be naturally distinguished in (\ref{integralV}) is the affine distance $\Delta v$, which is measured in the instantaneous rest frame of the accelerated observer. To study how short distances affect the particle spectrum seen by the accelerated observer, we restrict in (\ref{integral3}) the integration over $\Delta v$ to distances greater
than $\alpha\sim l_P \ll a^{-1}$. The result is
\be \label{expansion} - \frac{1}{4 \pi^2 w} \int_{|\Delta v |
> \alpha} d(\Delta v) \ e^{-i w\Delta v} \left[
\frac{(a/2)^2}{\sinh^2{\frac{a}{2}\Delta v}}-\frac{1}{(\Delta
v)^2}\right] \approx  \frac{1}{e^{2\pi w/a} -1}- \frac{\alpha a
}{12 \pi  w/a} + O(\alpha^3a^3)\ , \ee which shows that the
spectrum is not sensitive to a microscopic (Planckian)
cutoff\footnote{Note that this result is still valid even if the
two-point function is modified at short distances but the
principle of relativity, equivalence of
all inertial frames, is preserved (see Appendix).} $\alpha$ for $|\Delta v |$. \\

Let us now assume that equation (\ref{integralV}) is  referred to a particular inertial frame. This raises a problem, since the two-point function of the inertial observer, in the region $v\to -\infty$, would involve sub-Planckian distances, as discussed above in eq.(\ref{contraction}). One should, therefore, consider the effect of removing from (\ref{integralV}) the contribution of the two-point function of the inertial observer coming from sub-Planckian scales. In doing this, one sees that the particle spectrum turns out to be extremely sensitive to a microscopic cutoff for $|\Delta V |$, since then $|\Delta v|$ is macroscopic and much bigger than $a^{-1}$. In fact, for such $\Delta v$, the expansion (\ref{expansion}) is no longer valid, which casts doubts on the robustness of the Planckian spectrum. \\However, even if a short-distance cutoff is assumed for this (arbitrary) inertial observer, there is an additional argument supporting the robustness of the acceleration radiation. Instead of $H^+$, one can alternatively use the $H^-$ hypersurface for the calculation of the number of particles [see (\ref{integralU})]. Due to the existence of the completely
reflecting potential $V(\xi)\propto e^{2a\xi}\vec{k}_{\bot }^2$ for
the accelerated observer, the Rindler (wave-packet) modes with support
at $[v_1, v_2] \to -\infty$ have necessarily support at $[u_1, u_2]\to -\infty$
(see Fig. 1). In this situation we have instead
\be (U_1 - U_2)^2 \sim e^{-2au_1}(u_1 -u_2)^2 \ . \ee
\begin{figure}[htbp]
\begin{center}
\includegraphics[angle=0,width=3.6in,clip]{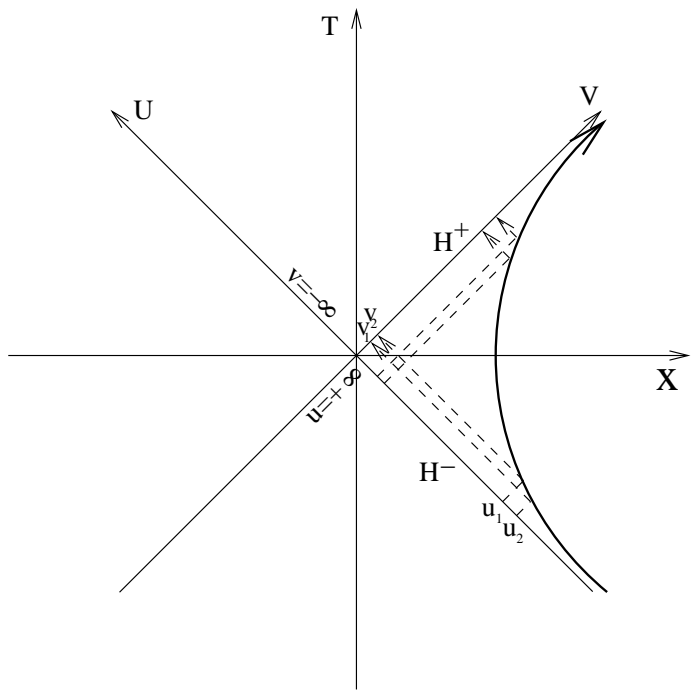}
\caption{A uniformly accelerating trajectory (bolded line). The
dotted lines represent the (Rindler) mode propagation from
$t=-\infty$ and $\xi = -\infty$, its reflexion by the potential
(around $\xi \sim 0$), until $t=+\infty$ and $\xi= -\infty$, as
described by the accelerating observer. }\label{Fig1}
\end{center}
\end{figure}
A Planckian cutoff $|U_1 - U_2|>\alpha\sim l_P$ for the inertial
affine distance in the region $[u_1, u_2] \to -\infty$  will now
remain sub-Planckian for the accelerated observer. One can,
therefore, restrict the integral in (\ref{integralU}) to distances
$|\Delta u|\ge \alpha$, which always imply $\Delta U>\alpha$, and
get an expression identical to (\ref{expansion}) without ever
invoking sub-Planckian scales. This shows that the
Fulling-Davies-Unruh effect can be  derived, in dimensions greater
than two, without ever invoking sub-Planckian distances (or extreme high energy scales).
This strongly suggests that the
acceleration radiation is indeed a low-energy phenomenon  and that it
should persist even if a Planck-length cutoff is introduced in the
theory. Note also that this reasoning cannot be used in black
holes due to the absence of a completely reflecting potential.
For the Hawking radiation, extra physical inputs are needed, as argued in
\cite{polchinski, agullo-navarro-salas-olmo-parker}.
It appears, therefore, that the acceleration radiation is, in any
case, more robust to trans-Planckian physics than Hawking radiation is.\\

It is important to note that if the above argument, namely the
interchangeable role of $H^+$ or $H^-$,  were not correct, as
applied to a modified theory with a Planck-length cutoff, one
would find a non-vanishing net flux of radiation. This can be seen
as follows. In the full relativistic theory (without any cutoff),
the accelerated observer perceives an energy flux to the right as
well as an (opposite) energy flux to the left. Summing up both
contributions the observer gets a null net flux of radiation but a
non-vanishing energy density. On physical grounds, this can be
seen as a consequence of  parity symmetry in our physical
scenario. If this symmetry is maintained in the presence of a
microscopic cutoff only a bath of radiation seems to be physically
acceptable.\footnote{Obviously, though no net flux of radiation is
allowed, there is a gradient of local temperature in the $X$
direction due to the redshift.}  We have seen that for the
calculation of the flux to the left (integration along $H^+$) at $v \to
-\infty$ ultrashort affine distances are required. However, for
the calculation of the flux to the right (integration along $H^-$) when $u
\to -\infty$, we do not need ultrashort affine distances to
generate the thermal spectrum. So both right and left fluxes
should be equal, and approximately thermal, to produce the bath of
radiation. The opposite argument applies at the trajectory points
$u \to +\infty$. In this case, ultrashort distances  are
apparently needed (for the inertial observer) in the computation
of the flux to the right, but not for the computation of the flux to the left.\\

In conclusion, we have pointed out that the acceleration radiation
effect can de rederived without actually invoking very high energy
physics. This supports the view that the  acceleration radiation is
robust against Planck scale physics and suggests that any theory
of quantum gravity, with new microscopic degrees of freedom,
should also reproduce this relativistic field-theory effect.  We
believe that our analysis of the acceleration radiation effect in
Minkowski space can also be extended to curved spacetimes. In
particular, for de Sitter space that would imply that the
semiclassical Gibbons-Hawking effect \cite{gibbons-hawking} would
remain robust against microscopic physics.

After completion of this work, we were informed that M.Rinaldi \cite{rinaldi} 
has recently reanalyzed the Unruh effect in terms of modified dispersion
relations. Conclusions similar to the present paper are also
displayed.

\section*{Appendix A}

Let us now illustrate the  discussion of the last section in terms
of a particle detector. To this end we shall modify the
relativistic theory by  deforming\footnote{Modifications via modified dispersion relations compatible with the principle of relativity have been studied in \cite{amelino-camelia-magueijo-smolin}.} the two-point function {\it ab
initio}. This has the advantage of going directly to the point of
interest to us, since it is just the form of the two-point
function that is relevant in the evaluation of the detector
response function. Obviously one could reconstruct the underlying
field theory to generate equations of motion, inner product, etc;
but all that is not necessary in the analysis below.

 For simplicity we shall considerer the case of a massless scalar field in four
dimensions. The simplest deformation of the two-point function is
\be \label{deformation2pointmassless}\langle \phi(x_1)\phi(x_2)
\rangle=  \frac{\hbar}{4\pi^2 (x_1 - x_2)^2 + l_P^2} \ . \ee The
rate ${\dot F(w)}$ can be now worked out, according to
(\ref{rateF}), as \be \label{naiveintegralrateF}{\dot F_{l_P}(w)}=
-\frac{\hbar}{4\pi^2}\int_{-\infty}^{+\infty}d\Delta \tau
e^{-iw\Delta \tau}  \frac{1}{\frac{4}{a^2}
\sinh^2(\frac{a}{2}\Delta \tau )+ l_P^2/4\pi^2}\ . \ee The result
is \be \frac{\hbar}{2\pi}\frac{ we^{\pi w/a}}{(e^{2\pi
w/a}-1)}\frac{\sinh[\frac{w}{a}(\theta-\pi)]}{\frac{w}{a}\sin\theta}
\ , \ee where $\theta\equiv 2\arcsin\left(\frac{l_P
a}{4\pi}\right)$. This largely departs from the thermal spectrum.
However it is not physically sound since, for an inertial observer
$a=0$, the response rate does not vanish, as one should expect
according to the principle of relativity. To produce a meaningful
expression one should subtract the naive ``inertial''
contribution, replacing (\ref{naiveintegralrateF}) by \be
\label{integralrateF}{\dot F_{l_P}(w)}=
-\frac{\hbar}{4\pi^2}\int_{-\infty}^{+\infty}d\Delta \tau
e^{-iw\Delta \tau}  \left[\frac{1}{
\frac{4}{a^2}\sinh^2(\frac{a}{2}\Delta \tau )+ l_P^2/4\pi^2} -
\frac{1}{(\Delta \tau)^2 + l_P^2/4\pi^2}\right]\ . \ee The final
result is then \be {\dot
F_{l_P}(w)}=\frac{\hbar}{2\pi}\left[\frac{we^{\pi w/a}}{(e^{2\pi
w/a}-1)}\frac{\sinh[\frac{w}{a}(\theta-\pi)]}{\frac{w}{a}\sin\theta}
+\frac{\pi e^{-wl_P/2\pi}}{l_P}\right]\ . \ee The thermal
Planckian spectrum is smoothly recovered in the limit
$\theta\approx l_P a/2\pi\to 0$. In fact, the rate ${\dot
F_{l_P}(w)}$ can be expanded as \be \label{eq:therm-alpha} {\dot
F_{l_P}(w)}\approx \frac{\hbar w}{2\pi}\left[\frac{1}{e^{2\pi
w/a}-1} - \frac{l_P a }{32\pi w/a}+ O(l_P^3 a^3)\right] \ . \ee
This result is in agreement with the estimation (\ref{expansion}).
Note that the crucial ingredient to preserve the thermal spectrum
is the requirement of having a vanishing detector response for all
inertial observers.

Thermality is maintained until a certain frequency scale $\Omega$,
which can be estimated by requiring positivity of
${\dot F_{l_P}(w)}$ in the above approximated expressions. A
simple calculation leads to the condition $32\pi\Omega e^{-2\pi
\Omega/a} \approx l_P a^2$. Planck-scale effects could potentially
emerge at the scale $\Omega$, which is roughly a few orders above
$T= a/2\pi $, if the acceleration is not very high.

\medskip

\acknowledgements

I.Agulló and G.J.Olmo thank MEC for FPU and postdoctoral grants,
respectively. I. Agulló thanks T. Jacobson, G.J. Olmo thanks W.
Unruh, and J.Navarro-Salas thanks A. Fabbri and M. Rinaldi for
useful discussions. This work has been partially supported by grants
FIS2005-05736-C03-03 and EU network MRTN-CT-2004-005104. G.J.Olmo
and L.Parker have been supported by NSF grants PHY-0071044 and
PHY-0503366. G.J.Olmo has been supported by Perimeter Institute for
Theoretical Physics. Research at Perimeter Institute is supported by
the Government of Canada through Industry Canada and by the Province
of Ontario through the Ministry of Research \& Innovation.

\end{document}